# Reference-State Error Mitigation: A Strategy for High Accuracy Quantum Computation of Chemistry

**Phalgun Lolur[1], Mårten Skogh[1,2], Christopher Warren[3], Janka Biznárová[3], Amr Osman[3], Giovanna Tancredi[3], Göran Wendin[3], Jonas Bylander[3] and Martin Rahm[1,*]**

[1] Department of Chemistry and Chemical Engineering, Chalmers University of Technology, Gothenburg, Sweden

[2] Data Science & Modelling, Pharmaceutical Science, R&D, AstraZeneca, Gothenburg, Sweden

[3] Department of Microtechnology and Nanoscience MC2, Chalmers University of Technology, Gothenburg, Sweden

* Corresponding author: martin.rahm@chalmers.se

**Abstract**

Decoherence and gate errors severely limit the capabilities of state-of-the-art quantum computers. This work introduces a strategy for reference-state error mitigation (REM) of quantum chemistry that can be straightforwardly implemented on current and near-term devices. REM can be applied alongside existing mitigation procedures, while requiring minimal post-processing and only one or no additional measurements. The approach is agnostic to the underlying quantum mechanical ansatz and is designed for the variational quantum eigensolver (VQE). Two orders-of-magnitude improvement in the computational accuracy of ground state energies of small molecules ($H_2$, $HeH^+$ and $LiH$) is demonstrated on superconducting quantum hardware. Simulations of noisy circuits with a depth exceeding 1000 two-qubit gates are used to argue for scalability of the method.

Quantum computers hold a potential for solving problems that are intractable on current and future computers.[1,2] Quantum chemistry is one of the research areas where *quantum advantage* is expected in the near future.[3–6] One of the major challenges in realizing practical quantum computation of chemistry is the sensitivity of quantum devices to noise. Errors due to noise can be caused by several factors such as spontaneous emission, control and measurement imperfection, and unwanted coupling with the environment.[7] Whereas reliable error correction is expected in future quantum computers, such *fault-tolerant* machines will put high demands on both quality and number of physical qubits.[4] Increasingly robust hybrid algorithms[2,8–10] are being designed for quantum chemistry on near-term, noisy intermediate-scale quantum (NISQ) devices.[11] Unfortunately, noise causes such algorithms to produce results, such as energies of molecules, that are of relatively low quality, even as they rely on shallow quantum circuits.[12–15] We will return to discuss how one can define quality in terms of accuracy and precision in this context.

The general challenge of noise in quantum hardware has motivated the development of several methods for *error mitigation*: readout/measurement error mitigation,[16] zero noise/Richardson extrapolation,[17,18] probabilistic error cancellation,[19] quantum subspace expansion,[20] virtual state distillation,[21] and symmetry verification[22] are some examples of techniques exploited to improve the quality of measurements of encoded Hamiltonians through pre- or post-processing. Such techniques have been shown to offer improvements when computing energies of small molecules with variational algorithms. Whereas these techniques are designed for shallow circuits and relatively low error-rates, they are generally not expected to work well with the deeper circuits required to encode more realistic problem sizes when subject to currently achievable error rates.[1]

In this study, we report on a chemistry-inspired error-mitigation strategy that can be combined with any variant of the variational quantum eigensolver[8,23] (VQE). Our approach, reference-state error mitigation (REM), relies on post-processing that can be readily performed on a classical computer. The method is applicable across a wide range of noise intensities and is low-cost in that it requires an overhead of at most one additional VQE energy evaluation. REM can readily be employed together with other error mitigation methods and throughout this work we additionally use readout mitigation, which corrects for hardware-specific non-ideal correlation between prepared states and measured states.[16]

**Reference-state Error Mitigation (REM)**

The goal of the VQE algorithm is to minimize the electronic energy with respect to a set of quantum circuit parameters, i.e.,

$$E_{VQE} = \min_{\vec{\theta}} E(\vec{\theta}) = \min_{\vec{\theta}} \langle \Psi(\vec{\theta})|\hat{H}|\Psi(\vec{\theta})\rangle, \qquad (1)$$

where $\vec{\theta} = [\theta_1, \theta_2, \dots \theta_n]$ and $\hat{H}$ is the molecular Hamiltonian. The VQE energy, $E_{VQE}(\vec{\theta})$, can therefore be thought of as living on an $n$-dimensional surface in parameter space. A one-dimensional representation of such a surface is shown in orange at the top of Figure 1. The $E_{VQE}(\vec{\theta})$ surface is associated with some degree of systematic and random noise that can only be partially removed by readout mitigation. Our theoretical discussion of errors below assumes that we measure the expectation value given by equation (1) exactly, i.e., assuming no random error in the evaluation. In practice we are limited to a finite number of measurements (samples), resulting in a spread in the measured energy values around the exact expectation value.

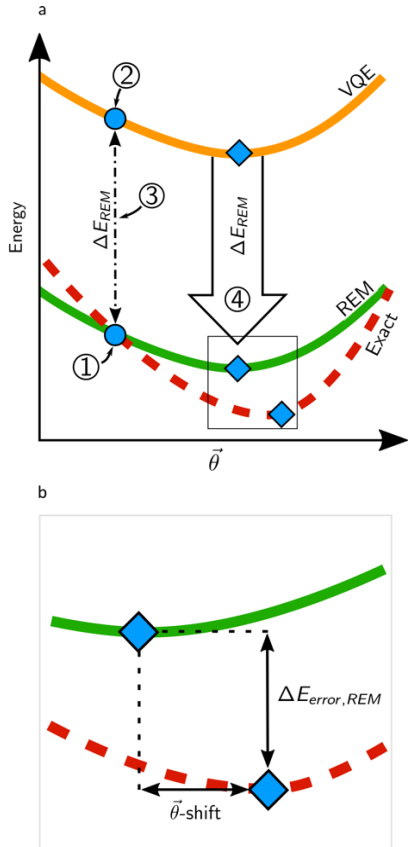

*Figure 1: (a) One-dimensional representation of the electronic energy E as a function of quantum circuit parameters $\vec{\theta}$. The REM approach can be explained as a four-step process: (1) A computationally tractable reference solution (such as Hartree-Fock) is computed on a classical computer. (2) A quantum measurement of the VQE surface (orange) is made using parameters that exactly correspond to the reference solution. (3) The difference in energy calculated for the reference solution on classical and quantum hardware defines an error, $\Delta E_{REM}$. (4) The error estimate is assumed to be systematic and is used to correct the VQE surface in the proximity of the reference coordinate. The resulting REM corrected surface and the exact (noise-free) solution are represented by solid green and dashed red lines, respectively. Blue dots and diamonds represent the coordinates of the reference calculations and minima of the different energy landscapes, respectively (b) Insert illustrates the how the error after application of REM is a function of the parameter dependence of noise.*

As the name suggests, the REM method rests on an appropriate choice of a reference wavefunction, or reference state. We recommend the reference state is a) chemically motivated, i.e., likely physically similar to the sought state, and b) fast (or at least viable) to evaluate using a classical computer. The Hartree-Fock state is a practical example of an often-suitable reference wavefunction that is based on a computationally efficient mean-field description of the electronic potential. We rely on Hartree-Fock in this work but note that other reference states can be used. Because an appropriate reference state usually also makes for a good initial guess for the VQE algorithm, it is often possible to perform REM without incurring *any* additional measurement cost.

Once the parameterized reference state $|\Psi(\vec{\theta}_{ref})\rangle$ is prepared, an *exact* determination of the resulting energy error $\Delta E_{REM}$ at the reference parameters can be made,

$$\Delta E_{REM} = E_{VQE}(\vec{\theta}_{ref}) - E_{exact}(\vec{\theta}_{ref}), \quad (2)$$

where $E_{exact}(\vec{\theta}_{ref})$ is the exact solution (up to numerical precision) for the reference state, evaluated on a classical computer. $E_{VQE}(\vec{\theta}_{ref})$ refers to the energy evaluated from measurements on a quantum computer at the reference parameter value, $\vec{\theta}_{ref}$. The exact energy at any arbitrary coordinate, $E_{exact}(\vec{\theta})$, can be expressed as,

$$E_{exact}(\vec{\theta}) = E_{VQE}(\vec{\theta}) - \Delta E_{REM} - \Delta E_p(\vec{\theta}), \quad (3)$$

where $\Delta E_p(\vec{\theta})$ includes any parameter-dependence of noise present, and $\Delta E_p(\vec{\theta}_{ref}) = 0$. The underlying assumption of the REM method is that such parameter dependence of the noise is negligible close to the reference geometry, i.e., $\lim_{\Delta\vec{\theta}\to 0} \Delta E_p(\vec{\theta}) = 0$ where $\Delta\vec{\theta} = |\vec{\theta} - \vec{\theta}_{ref}|$. In other words, the effectiveness of the REM approach can be assumed dependent on the Euclidean distance of the reference state to the exact solution $|\vec{\theta}_{exact} - \vec{\theta}_{ref}|$, given that both are in the same convex region of the energy surface. When this approximation fails, noise can shift features in the energy surface, such as the optimal coordinates identified using the VQE algorithm, $\vec{\theta}_{min,VQE}$, away from the true minimum, $\vec{\theta}_{min,exact}$ (Figure 1b). When evaluating our method, we will not quantify $\Delta E_p(\vec{\theta})$ but instead compare energies obtained for the two minima on the exact and the VQE surfaces, respectively,

$$\Delta E_{error,VQE} = E_{VQE}(\vec{\theta}_{min,VQE}) - E_{exact}(\vec{\theta}_{min,exact}). \quad (4)$$

In Eq. (4), $E_{exact}(\vec{\theta}_{min,exact})$ is the exact solution obtained by ideal noise-free VQE optimization and $E_{VQE}(\vec{\theta}_{min,VQE})$ is the energy of a converged noisy VQE optimization. The error remaining after applying REM to a converged noisy VQE optimization is,

$$\Delta E_{error,REM} = \Delta E_{REM} - \Delta E_{error,VQE}. \tag{5}$$

**Results and Discussion**

To assess the reliability of REM we have implemented it for VQE computation of small molecules on two current NISQ devices, the Quito of IBMQ and the Särimner of Chalmers. Details of hardware, circuits and measurements are provided in the SI. Table 1 shows an amalgamation of our measurement results for the ground state energy of the hydrogen molecule ($H_2$), helium hydride ($HeH^+$), and lithium hydride (LiH). The ansätze[3] used for the $H_2$ and $HeH^+$ molecules are chemistry-inspired and based on unitary coupled cluster theory, whereas a hardware-efficient ansatz is used for LiH. Table 1 also includes results of simulations of LiH and beryllium hydride ($BeH_2$). The latter circuits are substantially larger than what is feasible on current devices as they would incur insurmountable errors due to noise. Combined, this test set ranges from a 2-qubit circuit with just one two-qubit gate for $H_2$, to a 6-qubit circuit with 1096 two-qubit gates for $BeH_2$ (Table A1). Our simulations of $BeH_2$ contains 26 variational parameters (Table A1).

*Table 1: Total ground state energies of molecules at experimental equilibrium distances, without and with the application of REM. Readout mitigation has been applied for all the VQE calculations. All energies are given in hartree.*

| Molecule[a] | $E_{exact}(\vec{\theta}_{min})$ | $E_{VQE}(\vec{\theta}_{min,VQE})$ | $E_{REM}$ | $\Delta E_{error,VQE}$ | $\Delta E_{error,REM}$ |
|---|---|---|---|---|---|
| $H_2$[b] | -1.1373 | -1.1085 | -1.1355 | 0.0288 | 0.0018 |
| $HeH^+$[c] | -2.8542 | -2.8247 | -2.8544 | 0.0294 | -0.0002 |
| LiH[c] | -7.8787 | -7.6071 | -7.8651 | 0.2686 | 0.0136 |
| LiH[d] | -7.8811 | -7.3599 | -7.8705 | 0.5213 | 0.0106 |
| $BeH_2$[d] | -15.5895 | -13.9873 | -15.5632 | 1.6021 | 0.0263 |

*[a] Calculations refer to experimental bond distances from the National Institute of Standards and Technology (NIST). [b] Run on Chalmers Särimner with 5000 samples. [c] Run on IBMQ-Quito with 8192 samples. [d] Simulated results using a noise model from IBMQ-Athens.*

Table 1 shows how the application of REM brings down the error by approximately two orders of magnitude compared to regular VQE used together with readout mitigation. Without readout mitigation, VQE errors are substantially larger (Tables S3 and S6). The examples in Table 1 are sorted by increasing circuit depth (details of which are provided in Table S1) which together are suggestive of both robustness and a scalability of the approach. The remaining error after mitigation is consistently on the order of millihartree, and the magnitude of the REM correction grows with the complexity of the quantum circuit (*cf.* $H_2$ vs. $BeH_2$ in Table 1).

We note that REM consistently provides energies that are *lower* than those of the reference Hartree-Fock energies in our calculations, even at relatively high noise levels (Figure 3). In principle, unsuitable choices of reference states combined with significant parameter dependence of noise, $\Delta E_p(\vec{\theta}) \napprox 0$, might result in over-correction of the measured VQE energy, taking the energy below the true minimum, as can be seen in the results for $HeH^+$.

The major challenge when running VQE calculations with deep circuits on NISQ hardware is converging the optimization. This challenge of VQE is one of classical optimization, mostly due to vanishing gradients with increasing noise in a large parameter space. Our method does not help in this task since it conserves the shape of the energy landscape. However, REM does (given a proper reference state) guarantee a more computationally accurate measurement of the total energy upon VQE convergence.

Table 1 demonstrates the effectiveness of REM when applied to molecules in their equilibrium geometry, for which the degree of electron correlation is relatively low and the Hartree-Fock reference is suitable. Figure 2 shows how the REM method performs far from equilibrium for the bond dissociation of $H_2$ and HeH+. As expected, the effectiveness of our implementation of REM decreases somewhat in regions where the Hartree-Fock state offers a poor description, such as the stretched $H_2$ bond. However, the method consistently provides a substantial improvement across the potential energy surface. A more suitable description of the partially broken bond of $H_2$ should ideally account for the near degeneracy of multiple states, i.e., it would require a multi-reference wavefunction. The bond dissociation of HeH+ proceeds to He and an isolated proton, a state well described by a single-reference Hartree-Fock description. Figure 2 also illustrates the effect of readout mitigation,[16] which we use per default in all measurements and that we recommend together with REM. Other mitigation strategies may, in principle, also be combined with REM.

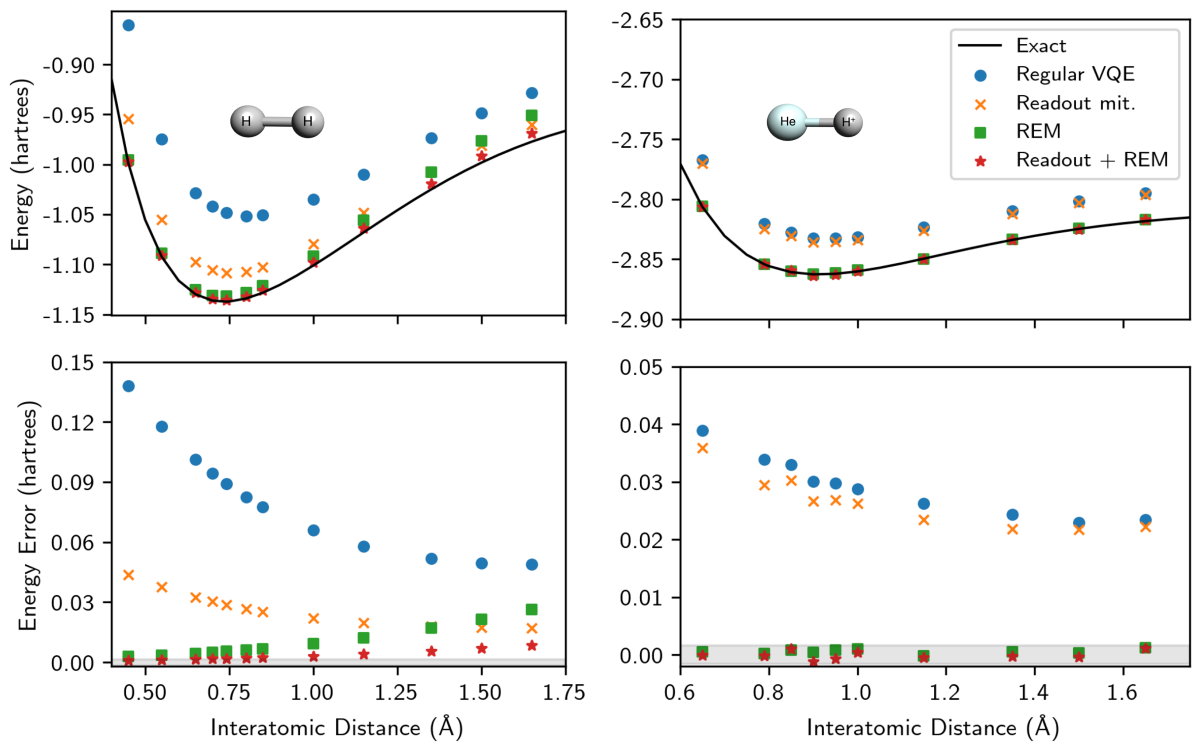

*Figure 2: Top: Potential energy surfaces for the dissociation of $H_2$ and HeH+. Exact noise-free solutions from state-vector simulations are represented by black lines. Regular VQE energies obtained using a quantum computer are shown as blue dots. Results following readout mitigation and REM are shown as orange crosses and green squares, respectively. The combination of both readout mitigation and REM is shown as red stars. Measurements for $H_2$ and HeH+ were performed on Chalmers Särimner and IBMQ-Quito, respectively. Bottom: error of the different approaches relative to the exact solution in the given minimal basis set. The gray region corresponds to an error of 1.6 millihartree (1 kcal/mol) with respect to the corresponding noise-free calculations.*

The robustness of REM was further evaluated by performing simulations of $H_2$ and HeH+ while varying the noise level. Noise was introduced in these simulations by modelling imperfect gate-fidelities as single ($S$) and two-qubit ($T$) depolarizing errors, connected through a linear relationship, $S = 0.1 * T$ (see the SI). This kind of noise modelling enables straightforward comparison with error rates on physical quantum devices (Figure 3). REM is shown to be effective despite the steady increase in single- and two-qubit depolarizing error rates, as expected.

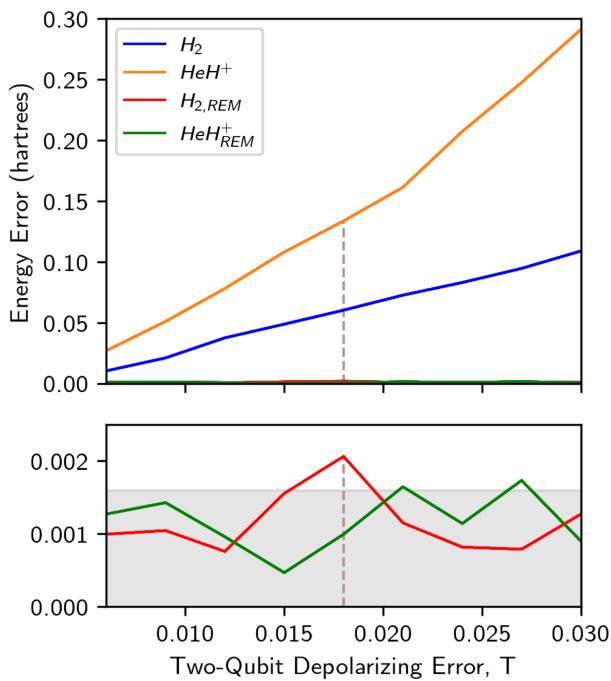

*Figure 3: Absolute errors as a function of increasing depolarizing errors for ground state calculations of $H_2$ and $HeH^+$. The two-qubit depolarizing error of the Chalmers Särimner device is indicated by a vertical line for reference. Bottom insert: Final energies following application of REM are largely independent of the noise level, and the computational accuracy is consistently close to or below 1.6 millihartree (1 kcal/mol) for these molecules.*

The evaluation of *computational accuracy* in this work should not be confused with *chemical accuracy*.[24] We here use the term *computational accuracy* specifically when comparing results of a quantum calculation with the exact solution *at that same level of theory*. Computational accuracy then, in the context of VQE calculations, refers to how exact a given VQE problem is solved with respect to the given Hamiltonian and ansatz. This accuracy can only be quantified so long as it is possible to solve the problem without noise, e.g., using conventional quantum chemistry (which we can still do; discussing the limits of conventional quantum chemistry methods is outside the scope of this work). Practical implementations of VQE on real hardware currently suffer from drastic deficiencies in level of theory, basis set size,[25] proper consideration of the physical environment (e.g., solvent effects), and dynamical effects. These limitations keep quantum computation (including our own) from accurately predicting real chemical processes. On the other hand, *chemical accuracy* is the correct term for what is required to make realistic predictions and is commonly defined as an error of 1 kcal/mol (~1.6 millihartree) from chemical experiments. No quantum computation has yet predicted molecular energies, reaction barrier heights, or the thermodynamics of a reaction to this accuracy. Despite this, there are several claims of "chemical accuracy" using NISQ devices.[8,13,26,27] We encourage the community to use the appropriate terminology. The hunt for chemical accuracy in the NISQ era is far from over.

## Conclusions

In this work we demonstrate an error mitigation strategy applicable to quantum chemical computations on NISQ devices. The REM method relies on determining the exact error in energy due to hardware and environmental noise for a reference wavefunction that can be feasibly evaluated on a classical computer. The most important underlying assumption of REM is negligible dependence of noise on circuit parameters in the vicinity of this reference wavefunction. In this work Hartree-Fock references are used, which describes the physics of the molecular states well enough for REM to perform effectively. The REM method is shown to drastically improve the computational accuracy at which total energies of molecules can be computed using current quantum hardware. How significant an improvement REM provides depends on the noise level. In our experiments, the computation accuracy is improved by two orders of magnitude using REM. With increasing noise, the procedure will return an energy that is equal to or below the classically evaluated reference. However, in the presence of substantial quantum circuit parameter dependence of noise it cannot be ruled out that REM may underestimate the true energy. The non-variational nature of error mitigation strategies remains a problem to solve. This mitigation strategy does not incur meaningful additional classical or quantum computational overhead and can be used to reduce errors on near-term devices by orders of magnitude when running VQE calculations. Because error rates vary both between NISQ devices and between circuits, it is not currently productive to rely on error cancellation, i.e., systematic errors inherent in quantum chemical levels of theory, when evaluating relative energies of chemical transformations. By enabling more precise evaluations of molecular total energies, REM moves us toward meaningful relative comparisons and toward chemical accuracy.

**Funding:** This research has been supported by funding from the Wallenberg Center for Quantum Technology (WACQT) and from the EU Flagship on Quantum Technology H2020-FETFLAG-2018-03 project 820363 OpenSuperQ.

**Acknowledgements:** This research relied on computational resources provided by the Swedish National Infrastructure for Computing (SNIC) at C3SE, NSC and PDC partially funded by the Swedish research council through grant agreement no 2018-05973. We also acknowledge experimental assistance from Christian Križan and Andreas Bengtsson as well as insights from Jorge Fernández Pendás at Chalmers University of Technology. The Chalmers device was made at Myfab Chalmers; it was packaged in a holder and printed circuit board with original designs shared by the Quantum Device Lab at ETH Zürich.

**Author contributions:**

Conceptualization: PL, MS, MR

Funding acquisition: GW, JBy

Investigation (Theoretical): PL, MS

Investigation (Experimental): CW, GT

Device Fabrication: JBi, AO

Supervision: MR, JBy, GW

Writing: MR, PL, MS, CW, JBy

**Competing Interests:** Authors declare that they have no competing interests.

# Appendix: Circuit complexity details

Table A1: Comparison of circuit complexities for $H_2$, $HeH^+$, LiH, and $BeH_2$ listed in order of increasing complexity. Details are provided in the SI.

| Molecule | Ansatz | Qubits | Depth | Two-Qubit Gates | Parameters |
|---|---|---|---|---|---|
| $H_2$ | Simplified UCCD | 2 | 5 | 1 | 1 |
| $HeH^+$ | UCCSD | 2 | 14 | 4 | 2 |
| LiH | Hardware-efficient | 4 | 9 | 6 | 8 |
| LiH | UCCSD | 4 | 275 | 172 | 8 |
| $BeH_2$ | UCCSD | 6 | 1480 | 1096 | 26 |

# Supporting Information
# Reference-State Error Mitigation: A Strategy for High Accuracy Quantum Computation of Chemistry

Phalgun Lolur[1], Mårten Skogh[1,2], Christopher Warren[3], Janka Biznárová[3], Amr Osman[3], Giovanna Tancredi[3], Göran Wendin[3], Jonas Bylander[3] and Martin Rahm[1,*]

1. Department of Chemistry and Chemical Engineering, Chalmers University of Technology, Gothenburg, Sweden
2. Data Science & Modelling, Pharmaceutical Science, R&D, AstraZeneca, Gothenburg, Sweden
3. Department of Microtechnology and Nanoscience MC2, Chalmers University of Technology, Gothenburg, Sweden

* Corresponding author: martin.rahm@chalmers.se

## Table of Content

## Notation

The following notation is used in the SI:

$E_{exact}(\vec{\theta}_{ref})$ – Energy of the reference state in the absence of noise

$E_{VQE}(\vec{\theta}_{ref})$ – Energy of the reference state from VQE

$E_{VQE*}(\vec{\theta}_{ref})$ – Energy of the reference state from VQE with readout mitigation

$E_{exact}(\vec{\theta}_{min})$ – Energy of the state of interest in the absence of noise

$E_{VQE}(\vec{\theta}_{min,VQE})$ – Energy of the state of interest from VQE

$E_{VQE*}(\vec{\theta}_{min,VQE})$ – Energy of the state of interest from VQE with readout mitigation

$E_{REM}$ – Energy of the state of interest from VQE with REM

$E_{REM*}$ – Energy of the state of interest from VQE with readout mitigation and REM

$\Delta E_{error,VQE}$ – Error of the VQE energy with respect to $E_{exact}(\vec{\theta}_{ref})$

$\Delta E_{error,VQE*}$ – Error of the readout mitigated VQE energy with respect to $E_{exact}(\vec{\theta}_{ref})$

$\Delta E_{error,REM}$ – Error of the REM mitigated VQE energy with respect to $E_{exact}(\vec{\theta}_{ref})$

$\Delta E_{error,REM*}$ – Error of the REM+readout mitigated VQE energy with respect to $E_{exact}(\vec{\theta}_{ref})$

$r$ – Interatomic distance (Å)

# Chalmers Device Details

We executed the quantum algorithm at Chalmers on a superconducting three-qubit quantum processor named Särimner, of which we only use two qubits ($Q_0$, $Q_1$). This device is shown in Figure S1 and consists of three transmon qubits,[1] coupled using a single tunable coupler, $C_1$. Single-qubit gates are implemented using on-chip drive lines to individually control each qubit with microwave pulses. Each qubit can be individually measured using its readout resonator with readout performed simultaneously using frequency multiplexed pulses on the common readout feedline. Two-qubit gates are activated via an AC flux-pulse applied to the coupler to modulate its frequency.[2] The coupler is itself a frequency-tunable transmon qubit, however, it only serves to mediate the interaction between pairs of qubits and itself never enters the computational space during operation. A full list of parameters of the device can be found in Table S1.

The AC flux-pulse which activates the coupling takes the form $\Phi(t) = \Phi_b + \Omega(t)\cos(\omega_d t)$, where $\Phi_b$ is the DC flux bias of the drive, $\Omega(t)$ is the envelope of the pulse which consists of a 25 ns cosine rise and fall and a 310 ns flat top, and $\omega_d$ is the carrier frequency. The carrier frequency is chosen such that it drives a controlled-Z (CZ) transition between $Q_0$ and $Q_1$. This is on resonance with the transition $|20\rangle \leftrightarrow |11\rangle$ and occurs at $\omega_d = |\omega_0 + \eta_0 - \omega_1|$. A full oscillation between these states brings about a conditional phase on the $|11\rangle$ state which can be calibrated to implement a CZ gate.

The parametric gate is useful in this higher connectivity architecture as the interaction between pairs can be selectively chosen. Different gates are activated between qubits when the frequencies between transitions for pairs of qubits are off-resonant from one another as they are in this device. The third qubit can also be detuned such that all transitions are far off-resonant and there is no risk of a frequency collision when operating the device with just $Q_0$ and $Q_1$.

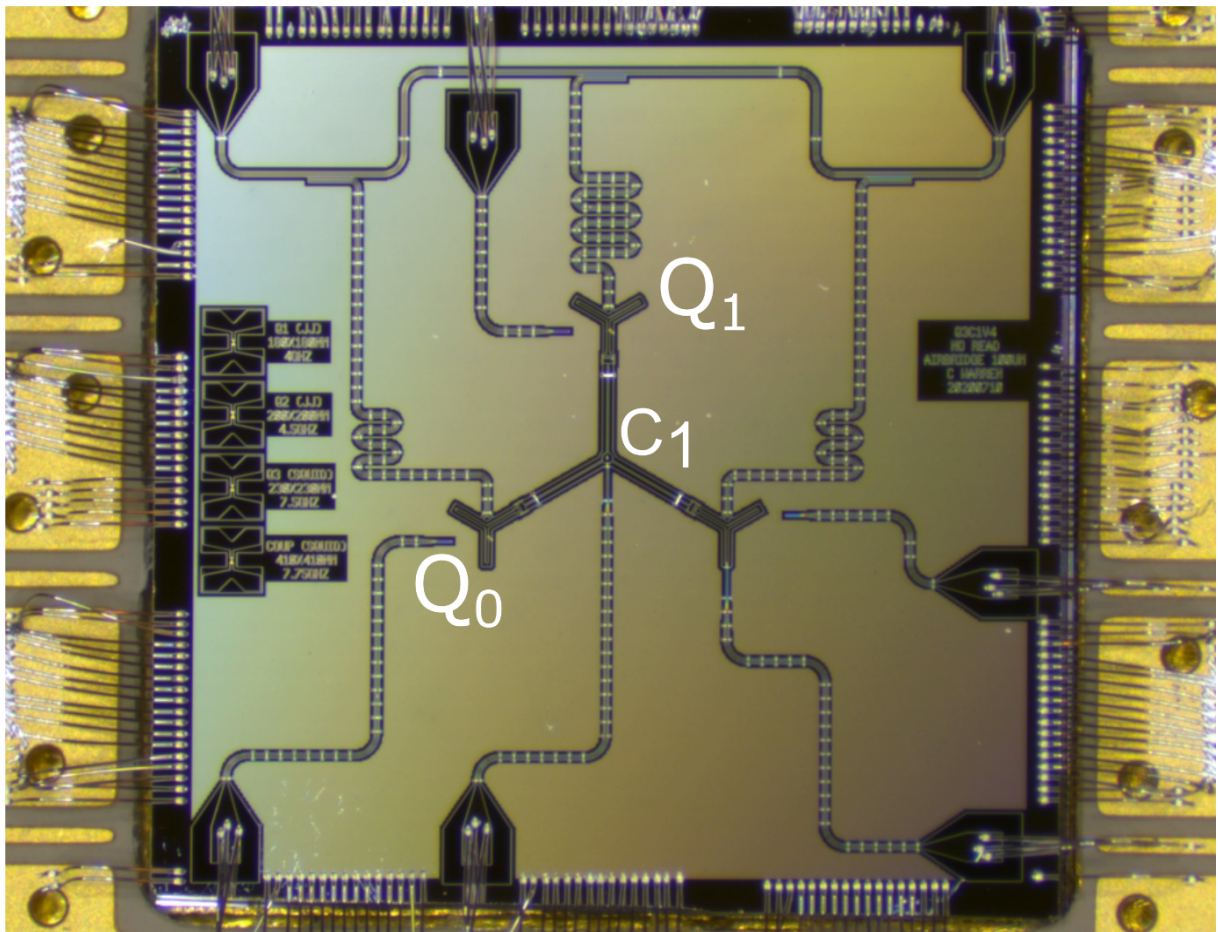

Figure S1: Microscope image of the bonded 3-qubit Särimner device from Chalmers. Only $Q_0$ and $Q_1$ are used in this work. The coupler, $C_1$, consists of a flux-tunable transmon qubit and serves to mediate interactions between pairs. Each qubit has individual readout and drive lines which are used to control the device. The readout lines are coupled to a shared feedline which is used for multiplexed readout.

Table S1: Experimental parameters for the 3-qubit Chalmers device Särimner. We report our average gate error achieved through randomized benchmarking for single qubit gates, as our gate set differs from that of IBM, as well as interleaved randomized benchmarking for our 2-qubit CZ gate.

| Qubit | $T_1$ [μs] | $T_2$ [μs] | Qubit Frequency, $\omega_i$ [GHz] | Anharmonicity, $\eta_i$ [GHz] | Single Qubit Gate Error | CZ-Gate Error |
|---|---|---|---|---|---|---|
| $Q_0$ | 35.98 | 38.74 | 3.799 | -0.1885 | 4.9e-4 | 1.8e-2 |
| $Q_1$ | 36.24 | 39.34 | 4.383 | -0.1837 | 5.2e-4 | 1.8e-2 |

The single- and two-qubit gate fidelities during the execution of the quantum algorithm were 99.95% and 98.2%. The device design, fabrication methods, measurement setup, gate implementation on hardware, and tune-up methods are described in detail elsewhere.[3]

**Readout Error Mitigation**

Readout error mitigation is performed by constructing a calibration of the confusion matrix $C$. The entries of this matrix, $C_{i,j}$, are the probabilities of preparing the state $|i\rangle$ and then measuring the state $|j\rangle$, i.e., $C_{i,j} = P(j | i)$. The matrix $C$ can then be used to correct a set of Pauli string measurements $\vec{m} = [P(0), ..., P(j), ..., P(n)]^T$ by either multiplying by the inverse confusion matrix, $C^{-1}\vec{m} = \vec{m}'$, performing a least-squares fit to reconstruct the most likely outcome, or by a procedure known as 'Bayesian Unfolding'.[4]

In this work, we mitigate our results by implementing a least-square fit to a quadratic cost function, where $\lambda(\vec{x}) = (\vec{m} - C\vec{x})^2$ with the constraint that the sum of the resulting vector must be 1 and each element itself is bounded in the interval [0,1]. This avoids some issues resulting from matrix inversion arising from small off-diagonal elements, which can lead to unphysical results. The Sequential Least Squares Programming[5] (SLSQP) optimizer was used to find the $\vec{x}$ that minimize the cost for each set of measured Pauli strings. The confusion matrix of the Chalmers Särimner device is reported in Figure S2.

Both options for readout mitigation are also available in Qiskit and can be straightforwardly implemented for calculations on real devices and simulations.

|  | Prepared State |  |  |  |
|---|---|---|---|---|
| Measured State | $|00\rangle$ | $|10\rangle$ | $|01\rangle$ | $|11\rangle$ |
| $|00\rangle$ | 96.8±0.21 | 5.9±0.59 | 5.9±0.67 | 0.4±0.08 |
| $|10\rangle$ | 1.1±0.12 | 92.1±0.57 | 0.1±0.03 | 5.6±0.56 |
| $|01\rangle$ | 2.0±0.15 | 0.1±0.04 | 93.0±0.69 | 5.7±0.58 |
| $|11\rangle$ | 0.0±0.01 | 1.9±0.17 | 1.1±0.13 | 88.4±0.86 |

Figure S2. Confusion matrix for the Chalmers Särimner device. The matrix consists of probabilities of measuring a state given a specific preparation. The confusion matrix is measured before each run of the VQE algorithm with 1000 shots and repeated 100 times to give an estimate for fluctuations in the readout.

# Hydrogen – $H_2$
## Ansatz, circuit and computation details

The $H_2$ wavefunction can be represented with four qubits, where each qubit corresponds to one molecular spin orbital in minimal STO-3G basis. Parity mapping[6] was here used to further reduce the problem by two qubits, utilizing the particle and spin conservation. PySCF[7] was used to generate the initial Hartree-Fock state. The circuit ansatz was constructed using a parameterized wavefunction based on unitary coupled cluster theory[6] as implemented in Qiskit 0.21.[8] Since single excitations do not contribute to the final ground state energy of $H_2$, only double excitations were included in the ansatz which reduced the complexity further, to a single parameter, $\theta$. The resulting circuit, compiled to Qiskit's native U1, U2 and U3 gates, is depicted in Figure S2.

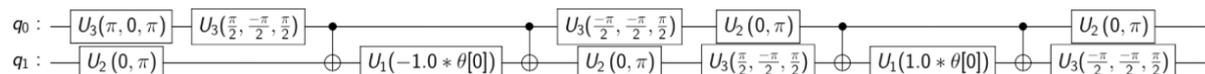

Figure S3: Quantum circuit for $H_2$ implementing double excitation of the unitary coupled-cluster operator with respect to the Hartree-Fock reference state. The circuit is compiled to the native gates of the IBMQ-Quito device.

The circuit shown in in Figure S2 was then shortened further by removing a repeated entangling step. A single entangling step was found sufficient to explore the Hilbert space of the problem. The resulting circuit was transpiled to gates native to the Särimner device (Figure S4) and is similar to the one used by Kandala et. Al.[9] The Särimner gate set consists of the set of single-qubit gates, $\left\{R_x(\pm\pi), R_y(\pm\pi), R_x\left(\pm\frac{\pi}{2}\right), R_y\left(\pm\frac{\pi}{2}\right), R_z(\theta)\right\}$, and two-qubit gate set {CZ}. Both circuits return the same energy up to the eighth decimal point of a hartree numerically, justifying our circuit design. Since the circuit contains only one parameter, it was varied in the interval [-π, π] to obtain energy as a function of the variational parameter at several geometries. The obtained energies were fit using the lmfit package[10] into a cosine function, $A\cos(\theta - \alpha)$, where $A$ and $\alpha$ are fit parameters, to give the minima at different geometries (Figure S5). All experimental calculations were run on the Chalmers Särimner device with 5000 shots. The details of the Hamiltonian operator to be minimized are given in Table S2 for different geometries of $H_2$. The exact solutions for the given circuit and Hamiltonians were calculated using QuTiP.[11]

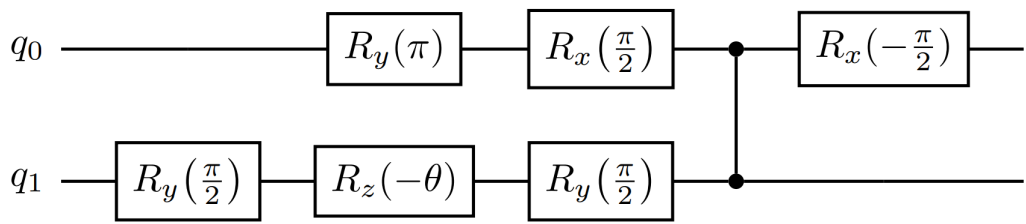

Figure S4: A compact quantum circuit for $H_2$ implementing double excitation of the coupled-cluster operator with respect to the Hartree-Fock reference state. The circuit is compiled to the native gates of Chalmers Särimner device.

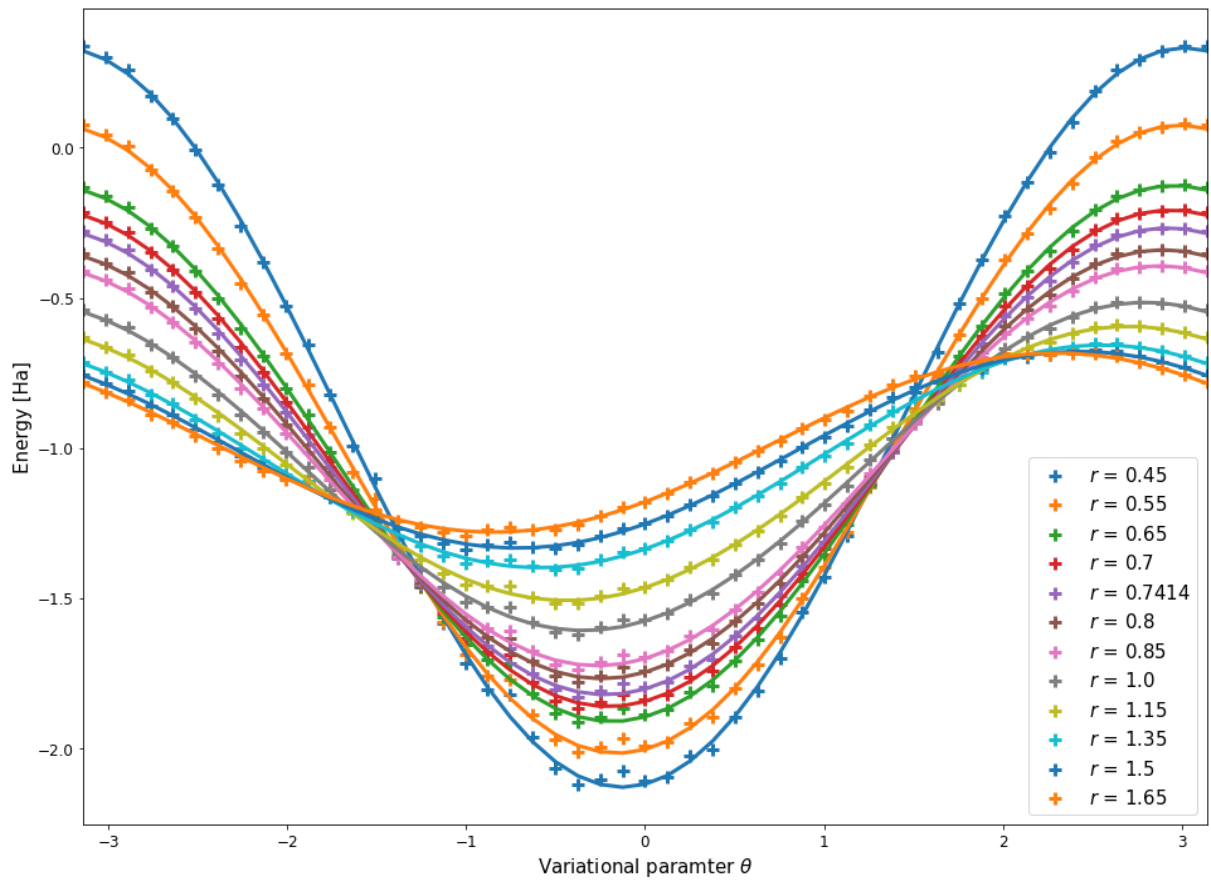

Figure S5: Measurement results for a single sweep of $\theta$ between $-\pi$ and $\pi$. The same measurement results have been used with different electronic Hamiltonians to generate the energies using QuTiP[11] for the different geometries. A cosine function of the form, $A\cos(\theta - \alpha)$, is used to generate the fits for the data.

## H₂ Hamiltonian

Table S2: The electronic Hamiltonian for H$_2$ expressed as weighted Pauli string operators after parity mapping for various geometry.

| r | II | IZ | ZI | ZZ | XX |
|---|---|---|---|---|---|
| 0.45 | -0.908 | 0.634 | -0.634 | -0.013 | 0.167 |
| 0.55 | -0.981 | 0.536 | -0.536 | -0.012 | 0.171 |
| 0.65 | -1.028 | 0.455 | -0.455 | -0.012 | 0.176 |
| 0.70 | -1.044 | 0.420 | -0.420 | -0.012 | 0.179 |
| 0.7414 | -1.054 | 0.394 | -0.394 | -0.011 | 0.181 |
| 0.80 | -1.063 | 0.360 | -0.360 | -0.011 | 0.185 |
| 0.85 | -1.068 | 0.334 | -0.334 | -0.010 | 0.188 |
| 1.00 | -1.069 | 0.268 | -0.268 | -0.009 | 0.197 |
| 1.15 | -1.058 | 0.215 | -0.215 | -0.007 | 0.206 |
| 1.35 | -1.033 | 0.161 | -0.161 | -0.005 | 0.220 |
| 1.50 | -1.010 | 0.129 | -0.129 | -0.004 | 0.230 |
| 1.65 | -0.985 | 0.103 | -0.103 | -0.003 | 0.239 |

## Data

The total energy of a system is the sum of its electronic and nuclear repulsion energies at a given geometry for $H_2$. The optimal parameters are reported in Table S3. The nuclear energies are reported in Table S4, and the electronic energies are reported in Table S5.

Table S3: Optimized parameters, before and after readout mitigation, that minimize the Hamiltonian operator on the Chalmer Särimner device for calculations of $H_2$.

| $r$ | $\theta$ (Uncorrected) | $\theta$ (Readout mit.) |
|---|---|---|
| 0.45 | -0.1186 | -0.1272 |
| 0.55 | -0.1437 | -0.1540 |
| 0.65 | -0.1737 | -0.1861 |
| 0.70 | -0.1906 | -0.2042 |
| 0.7414 | -0.2056 | -0.2202 |
| 0.80 | -0.2284 | -0.2445 |
| 0.85 | -0.2495 | -0.2669 |
| 1.00 | -0.3220 | -0.3438 |
| 1.15 | -0.4106 | -0.4372 |
| 1.35 | -0.5553 | -0.5876 |
| 1.50 | -0.6802 | -0.7153 |
| 1.65 | -0.8121 | -0.8477 |

Table S4: Nuclear repulsion energies ($V_{NN}$) of $H_2$ at various geometries, $r$. All $V_{NN}$ energies in hartrees.

| $r$ | $V_{NN}$ |
|---|---|
| 0.45 | 1.1759 |
| 0.55 | 0.9621 |
| 0.65 | 0.8141 |
| 0.70 | 0.7560 |
| 0.7414 | 0.7138 |
| 0.80 | 0.6615 |
| 0.85 | 0.6226 |
| 1.00 | 0.5292 |
| 1.15 | 0.4602 |
| 1.35 | 0.3920 |
| 1.50 | 0.3528 |
| 1.65 | 0.3207 |

Table S5: Exact, regular VQE and mitigated electronic energies and errors of H$_2$ at various geometries, $r$. All energies are in hartrees.

| $r$ [Å] | $E_{exact}(\vec{\theta}_{ref})$ | $E_{VQE}(\vec{\theta}_{ref})$ | $E_{VQE*}(\vec{\theta}_{ref})$ | $E_{exact}(\vec{\theta}_{min})$ | $E_{VQE}(\vec{\theta}_{min,VQE})$ | $E_{VQE*}(\vec{\theta}_{min,VQE})$ | $E_{REM}$ | $E_{REM*}$ | $\Delta E_{error,VQE}$ | $\Delta E_{error,VQE*}$ | $\Delta E_{error,REM}$ | $\Delta E_{error,REM*}$ |
|---|---|---|---|---|---|---|---|---|---|---|---|---|
| 0.45 | -0.9875 | -0.8524 | -0.9446 | -0.9984 | -0.8604 | -0.9546 | -0.9955 | -0.9975 | 0.1380 | 0.0438 | 0.0029 | 0.0010 |
| 0.55 | -1.0791 | -0.9649 | -1.0426 | -1.0926 | -0.9749 | -1.0550 | -1.0890 | -1.0914 | 0.1178 | 0.0376 | 0.0036 | 0.0012 |
| 0.65 | -1.1130 | -1.0162 | -1.0820 | -1.1299 | -1.0287 | -1.0974 | -1.1254 | -1.1284 | 0.1013 | 0.0325 | 0.0045 | 0.0015 |
| 0.70 | -1.1173 | -1.0281 | -1.0886 | -1.1362 | -1.0419 | -1.1058 | -1.1312 | -1.1345 | 0.0943 | 0.0304 | 0.0050 | 0.0017 |
| 0.7414 | -1.1167 | -1.0331 | -1.0897 | -1.1373 | -1.0482 | -1.1085 | -1.1318 | -1.1355 | 0.0891 | 0.0288 | 0.0055 | 0.0018 |
| 0.80 | -1.1109 | -1.0345 | -1.0861 | -1.1341 | -1.0516 | -1.1073 | -1.1280 | -1.1321 | 0.0825 | 0.0268 | 0.0062 | 0.0020 |
| 0.85 | -1.1025 | -1.0317 | -1.0794 | -1.1284 | -1.0507 | -1.1030 | -1.1215 | -1.1261 | 0.0776 | 0.0253 | 0.0068 | 0.0023 |
| 1.00 | -1.0661 | -1.0093 | -1.0473 | -1.1012 | -1.0352 | -1.0793 | -1.0919 | -1.0981 | 0.0660 | 0.0219 | 0.0092 | 0.0030 |
| 1.15 | -1.0210 | -0.9752 | -1.0054 | -1.0679 | -1.0099 | -1.0483 | -1.0557 | -1.0639 | 0.0580 | 0.0196 | 0.0122 | 0.0040 |
| 1.35 | -0.9572 | -0.9227 | -0.9449 | -1.0251 | -0.9733 | -1.0072 | -1.0078 | -1.0195 | 0.0517 | 0.0179 | 0.0172 | 0.0056 |
| 1.50 | -0.9109 | -0.8830 | -0.9005 | -0.9981 | -0.9486 | -0.9808 | -0.9765 | -0.9912 | 0.0495 | 0.0173 | 0.0216 | 0.0070 |
| 1.65 | -0.8678 | -0.8452 | -0.8590 | -0.9771 | -0.9283 | -0.9599 | -0.9508 | -0.9688 | 0.0489 | 0.0172 | 0.0263 | 0.0084 |

# Helium hydride – HeH$^+$

## Ansatz, circuit and computation details

Similar to H$_2$, the HeH+ wavefunction can be represented with two qubits using parity mapping[6] in a minimal basis. PySCF[7] was used to generate the initial Hartree-Fock state and the circuit ansatz was constructed as a parameterized wavefunction based on unitary coupled cluster theory[6] as implemented in Qiskit 0.21.[8] The circuit consists of three parameters – two single excitation parameters, *θ[0]* and *θ[1]*, and a double excitation parameter, *θ[2]*. These parameters are optimized using the VQE algorithm implemented in Qiskit and run on the IBMQ-Quito device with 8192 shots. The details of the Hamiltonian operator to be minimized are given in Table S6 for different geometries of HeH$^+$.

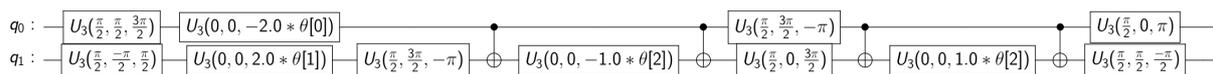

Figure S6: Quantum circuit for HeH$^+$ implementing the UCCSD operator with respect to the Hartree-Fock reference state. The circuit is compiled to the native gates of IBMQ-Quito device.

# HeH⁺ Hamiltonian

The details of the Hamiltonian operator to be minimized are given in Table S6 for different geometries of HeH⁺.

Table S6: The electronic Hamiltonian expressed as weighted Pauli string operators after parity mapping for various geometry of HeH⁺.

| $r$ | II | IZ | ZI | ZZ | ZX | XZ | IX | XI | XX |
|---|---|---|---|---|---|---|---|---|---|
| 0.65 | -3.229 | 0.635 | -0.635 | -0.074 | -0.094 | 0.094 | 0.094 | 0.094 | 0.157 |
| 0.7899 | -3.161 | 0.560 | -0.560 | -0.097 | -0.106 | 0.106 | 0.106 | 0.106 | 0.144 |
| 0.85 | -3.129 | 0.538 | -0.538 | -0.108 | -0.111 | 0.111 | 0.111 | 0.111 | 0.137 |
| 0.90 | -3.101 | 0.523 | -0.523 | -0.118 | -0.114 | 0.114 | 0.114 | 0.114 | 0.131 |
| 0.95 | -3.073 | 0.512 | -0.512 | -0.128 | -0.117 | 0.117 | 0.117 | 0.117 | 0.124 |
| 1.00 | -3.045 | 0.503 | -0.503 | -0.139 | -0.119 | 0.119 | 0.119 | 0.119 | 0.117 |
| 1.15 | -2.962 | 0.488 | -0.488 | -0.173 | -0.122 | 0.122 | 0.122 | 0.122 | 0.095 |
| 1.35 | -2.857 | 0.488 | -0.488 | -0.217 | -0.115 | 0.115 | 0.115 | 0.115 | 0.066 |
| 1.5 | -2.785 | 0.495 | -0.495 | -0.247 | -0.104 | 0.104 | 0.104 | 0.104 | 0.047 |
| 1.65 | -2.721 | 0.506 | -0.506 | -0.273 | -0.090 | 0.090 | 0.090 | 0.090 | 0.032 |

## Data

The total energy of a system is the sum of its electronic and nuclear repulsion energies at a given geometry for H₂. The optimal parameters are reported in Table S7. The nuclear energies are reported in Table S8, and the electronic energies are reported in Table S9.

Table S7: The optimized parameter values that minimize the Hamiltonian operator on the IBMQ-Quito device for various geometry of HeH$^+$. The angles are in radians.

| $r$ | $\theta[0]$ | $\theta[1]$ | $\theta[2]$ |
|---|---|---|---|
| 0.65 | 0.011 | 0.008 | -0.061 |
| 0.7899 | 0.014 | 0.016 | -0.067 |
| 0.85 | 0.013 | 0.010 | -0.065 |
| 0.90 | 0.012 | 0.013 | -0.063 |
| 0.95 | 0.017 | 0.015 | -0.065 |
| 1.00 | 0.021 | 0.021 | -0.063 |
| 1.15 | 0.017 | 0.017 | -0.053 |
| 1.35 | 0.012 | 0.012 | -0.036 |
| 1.5 | 0.009 | 0.003 | -0.025 |
| 1.65 | 0.008 | 0.005 | -0.018 |

Table S8: Nuclear repulsion energies ($V_{NN}$) of HeH$^+$ at various geometries, $r$. All $V_{NN}$ energies in hartrees.

| $r$ | $V_{NN}$ |
|---|---|
| 0.65 | 1.6282 |
| 0.7899 | 1.3399 |
| 0.85 | 1.2451 |
| 0.90 | 1.1759 |
| 0.95 | 1.1141 |
| 1.00 | 1.0584 |
| 1.15 | 0.9203 |
| 1.35 | 0.7840 |
| 1.5 | 0.7056 |
| 1.65 | 0.6414 |

Table S9: The exact, uncorrected and mitigated electronic energies and errors of HeH$^+$ at various geometries, $r$. All energies are in hartrees.

| $r$ [Å] | $E_{exact}(\vec{\theta}_{ref})$ | $E_{VQE}(\vec{\theta}_{ref})$ | $E_{VQE*}(\vec{\theta}_{ref})$ | $E_{exact}(\vec{\theta}_{min})$ | $E_{VQE}(\vec{\theta}_{min,VQE})$ | $E_{VQE*}(\vec{\theta}_{min,VQE})$ | $E_{REM}$ | $E_{REM*}$ | $\Delta E_{error,VQE}$ | $\Delta E_{error,VQE*}$ | $\Delta E_{error,REM}$ | $\Delta E_{error,REM*}$ |
|---|---|---|---|---|---|---|---|---|---|---|---|---|
| 0.65 | -2.7964 | -2.7580 | -2.7604 | -2.8062 | -2.7673 | -2.7703 | -2.8057 | -2.8063 | 0.0389 | 0.0359 | 0.0005 | -0.0001 |
| 0.7899 | -2.8447 | -2.8110 | -2.8150 | -2.8542 | -2.8203 | -2.8247 | -2.8540 | -2.8544 | 0.0338 | 0.0294 | 0.0002 | -0.0002 |
| 0.85 | -2.8517 | -2.8195 | -2.8225 | -2.8608 | -2.8278 | -2.8305 | -2.8600 | -2.8597 | 0.0330 | 0.0302 | 0.0008 | 0.0010 |
| 0.90 | -2.8540 | -2.8244 | -2.8261 | -2.8626 | -2.8326 | -2.8359 | -2.8622 | -2.8638 | 0.0300 | 0.0267 | 0.0004 | -0.0012 |
| 0.95 | -2.8542 | -2.8253 | -2.8267 | -2.8622 | -2.8324 | -2.8353 | -2.8614 | -2.8629 | 0.0298 | 0.0269 | 0.0008 | -0.0007 |
| 1.00 | -2.8529 | -2.8252 | -2.8270 | -2.8602 | -2.8315 | -2.8339 | -2.8592 | -2.8598 | 0.0287 | 0.0263 | 0.0010 | 0.0004 |
| 1.15 | -2.8445 | -2.8181 | -2.8206 | -2.8495 | -2.8233 | -2.8261 | -2.8497 | -2.8500 | 0.0262 | 0.0235 | -0.0002 | -0.0004 |
| 1.35 | -2.8314 | -2.8076 | -2.8093 | -2.8339 | -2.8095 | -2.8120 | -2.8333 | -2.8341 | 0.0243 | 0.0219 | 0.0005 | -0.0003 |
| 1.5 | -2.8234 | -2.8008 | -2.8013 | -2.8247 | -2.8017 | -2.8029 | -2.8244 | -2.8251 | 0.0230 | 0.0218 | 0.0003 | -0.0004 |

# Lithium hydride – LiH

## Ansatz, circuit and computation details

A LiH wavefunction can be represented with twelve qubits where each qubit corresponds to one molecular spin orbital in a minimal STO-3G basis. Symmetries in parity mapping[6] was used to further reduce the problem size by two qubits to ten qubits. As shown by Kandala *et al*[9], removing the orbitals that do not participate in bonding can bring down the problem size by four qubits. Coupled with the frozen core approximation, the final LiH circuit can be represented by just four qubits in and around equilibrium geometry. PySCF[7] was used to generate the initial Hartree-Fock state. A hardware efficient ansatz, inspired by Qiskit's two-local circuit class is used to construct a compact circuit representing LiH. It consists of alternating rotating layers of entanglement layers and is chosen to utilize the connectivity of IBMQ-Quito's connectivity as shown in Figure S7.

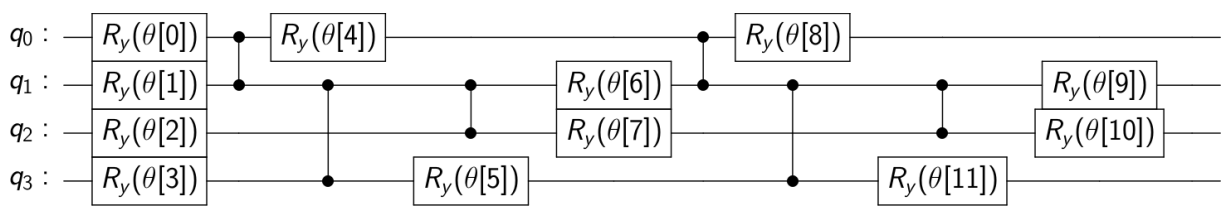

Figure S7: Quantum circuit for LiH utilizing a hardware-efficient ansatz.

# LiH Hamiltonian details at 1.5949 Å

Table S10: The electronic Hamiltonian of LiH can be expressed as weighted Pauli string operators with the following coefficients:

| | | | | | | | |
|---|---|---|---|---|---|---|---|
| IIII | -0.207 | ZXIZ | 0.012 | XZIZ | -0.013 | YYXZ | 0.008 |
| IIIZ | -0.094 | IXZX | -0.003 | XIZX | -0.002 | XXXZ | -0.008 |
| IIZX | -0.003 | ZXZX | -0.003 | XZZX | 0.002 | YYXI | 0.008 |
| IIIX | 0.003 | IXIX | 0.003 | XIIX | 0.002 | XXXI | -0.008 |
| IIXX | -0.001 | ZXIX | 0.003 | XZIX | -0.002 | ZZZZ | 0.084 |
| IIYY | 0.001 | IXXX | -0.009 | XIXX | -0.008 | ZZXZ | -0.009 |
| IIZZ | -0.212 | ZXXX | -0.009 | XZXX | 0.008 | ZZXI | -0.009 |
| IIXZ | 0.019 | IXYY | 0.009 | XIYY | 0.008 | XIZZ | -0.009 |
| IIXI | 0.019 | ZXYY | 0.009 | XZYY | -0.008 | XZZZ | 0.009 |
| IIZI | 0.359 | YYIZ | 0.032 | ZIIZ | 0.114 | XIXZ | 0.007 |
| IZII | 0.094 | XXIZ | -0.032 | ZIZX | -0.011 | XZXZ | -0.007 |
| ZXII | 0.003 | YYZX | -0.009 | ZIIX | 0.011 | XIXI | 0.007 |
| IXII | 0.003 | XXZX | 0.009 | ZIXX | -0.034 | XZXI | -0.007 |
| XXII | -0.001 | YYIX | 0.009 | ZIYY | 0.034 | ZIZZ | 0.060 |
| YYII | 0.001 | XXIX | -0.009 | IZZZ | -0.056 | ZIXZ | 0.011 |
| ZZII | -0.212 | YYXX | -0.031 | IZXZ | -0.013 | ZIXI | 0.011 |
| XZII | -0.019 | XXXX | 0.031 | IZXI | -0.013 | IZZI | 0.114 |
| XIII | 0.019 | YYYY | 0.031 | IXZZ | -0.002 | IXZI | -0.011 |
| ZIII | -0.359 | XXYY | -0.031 | ZXZZ | -0.002 | ZXZI | -0.011 |
| IZIZ | -0.122 | ZZIZ | 0.056 | IXXZ | 0.002 | YYZI | -0.034 |
| IZZX | 0.012 | ZZZX | 0.002 | ZXXZ | 0.002 | XXZI | 0.034 |
| IZIX | -0.012 | ZZIX | -0.002 | IXXI | 0.002 | ZZZI | -0.060 |
| IZXX | 0.032 | ZZXX | 0.003 | ZXXI | 0.002 | XIZI | -0.011 |
| IZYY | -0.032 | ZZYY | -0.003 | YYZZ | -0.003 | XZZI | 0.011 |
| IXIZ | 0.012 | XIIZ | 0.013 | XXZZ | 0.003 | ZIZI | -0.113 |

## Data

Frozen core energy: -7.7983328 hartrees

Nuclear repulsion energy: 0.99538004 hartrees

Table S11: The exact, uncorrected, and mitigated electronic energies and errors of LiH at 1.5949 Å. All energies are in hartrees.

| $E_{exact}(\vec{\theta}_{ref})$ | $E_{VQE}(\vec{\theta}_{ref})$ | $E_{VQE*}(\vec{\theta}_{ref})$ | $E_{exact}(\vec{\theta}_{min})$ | $E_{VQE}(\vec{\theta}_{min,VQE})$ | $E_{VQE*}(\vec{\theta}_{min,VQE})$ | $E_{REM}$ | $E_{REM*}$ | $\Delta E_{error,VQE}$ | $\Delta E_{error,VQE*}$ | $\Delta E_{error,REM}$ | $\Delta E_{error,REM}$ |
|---|---|---|---|---|---|---|---|---|---|---|---|
| -7.8620 | -7.6064 | -7.6071 | -7.8787 | -7.6071 | -7.6102 | -7.8627 | -7.8651 | 0.2717 | 0.2686 | 0.0160 | 0.0136 |

Optimal Parameters:

Table S12: The optimized parameter values that minimize the Hamiltonian operator on the IBMQ-Quito device for LiH at 1.5949 Å. The angles are in radians.

| $\theta[0]$ | $\theta[1]$ | $\theta[2]$ | $\theta[3]$ | $\theta[4]$ | $\theta[5]$ | $\theta[6]$ | $\theta[7]$ | $\theta[8]$ | $\theta[9]$ | $\theta[10]$ | $\theta[11]$ |
|---|---|---|---|---|---|---|---|---|---|---|---|
| 3.8987 | -6.5469 | -1.2442 | -5.0653 | 1.5509 | 2.0379 | 3.1205 | -4.7523 | 2.3617 | 6.2591 | -5.9394 | 3.2559 |

## Simulation Details for Lithium Hydride (LiH) and Beryllium Hydride (BeH$_2$)

As shown in the previous section, LiH wavefunction can be represented with four qubits. Making similar approximations as shown by Kandala *et al*[9], the BeH$_2$ wavefunction can be represented by six qubits when using the frozen core approximation and orbital reduction. PySCF[7] was used to generate the initial Hartree-Fock state. Qiskit's UCCSD module is used to construct the problem ansatz. The circuits are too large to be represented here but the details of the circuit can be found in Table A1. A noise-model from IBMQ-Athens has been added to the simulations to replicate real-world noisy behavior. All simulations have been run using 20,000 shots and repeated 5 times to ensure a high number of samples. The sampling noise is expressed in terms of the standard deviations of our errors. Readout mitigation has been applied for all the simulations.

*Table S13: Total ground state energies of molecules(simulated) at experimental equilibrium distances, without and with the application of REM. Bond distances have been obtained from the National Institute of Standards and Technology (NIST). A noise model from IBMQ-Athens has been added to all the simulations. Readout mitigation has been applied for all the VQE calculations. All energies are given in Hartree. The sampling error of the simulations is represented as the standard deviation.*

| Molecule | $E_{exact}(\vec{\theta}_{min})$ | $E_{VQE}*(\vec{\theta}_{min,VQE})$ | $E_{REM*}$ | $\Delta E_{error,VQE}*$ | $\Delta E_{error,REM*}$ |
|---|---|---|---|---|---|
| LiH | -7.8811 | -7.3599 | -7.8705 | 0.5213 ± 0.003 | 0.0106 ± 0.002 |
| BeH$_2$ | -15.5895 | -13.9873 | -15.5632 | 1.6021 ± 0.005 | 0.0263 ± 0.007 |

# IBM-Quito Device calibration and connectivity details

Table S13: IBM-Quito's calibration details, as imported from IBM Quantum Services. The device reports a quantum volume of 32.

| Qubit | T1 (μs) | T2 (μs) | Frequency (GHz) | Anharmonicity (GHz) | Readout assignment error | Prob meas0 prep1 | Prob meas1 prep0 | Readout length (ns) | ID error | Single-qubit Pauli-X error | CNOT error | Gate time (ns) |
|---|---|---|---|---|---|---|---|---|---|---|---|---|
| Q0 | 85.96 | 95.63 | 5.300 | -0.33148 | 0.0300 | 0.0468 | 0.0132 | 5351.111 | $2.93 \cdot 10^{-4}$ | $2.93 \cdot 10^{-4}$ | 0_1:$6.276 \cdot 10^{-3}$ | 0_1:234.667 |
| Q1 | 125.84 | 128.97 | 5.081 | -0.31925 | 0.0134 | 0.0208 | 0.0060 | 5351.111 | $2.92 \cdot 10^{-4}$ | $2.92 \cdot 10^{-4}$ | 1_3:$1.079 \cdot 10^{-2}$; 1_2:$7.076 \cdot 10^{-3}$; 1_0:$6.276 \cdot 10^{-3}$ | 1_3:334.222; 1_2:298.667; 1_0:270.222 |
| Q2 | 81.75 | 123.32 | 5.322 | -0.33232 | 0.0237 | 0.0358 | 0.0116 | 5351.111 | $2.55 \cdot 10^{-4}$ | $2.55 \cdot 10^{-4}$ | 2_1:$7.076 \cdot 10^{-3}$ | 2_1:263.111 |
| Q3 | 96.32 | 10.29 | 5.164 | -0.33508 | 0.0281 | 0.0454 | 0.0108 | 5351.111 | $3.13 \cdot 10^{-4}$ | $3.13 \cdot 10^{-4}$ | 3_4:$1.557 \cdot 10^{-3}$; 3_1:$1.079 \cdot 10^{-2}$ | 3_4:277.333; 3_1:369.778 |
| Q4 | 148.13 | 162.06 | 5.052 | -0.31926 | 0.0208 | 0.0328 | 0.0088 | 5351.111 | $2.74 \cdot 10^{-4}$ | $2.74 \cdot 10^{-4}$ | 4_3:$1.557 \cdot 10^{-2}$ | 4_3:312.889 |

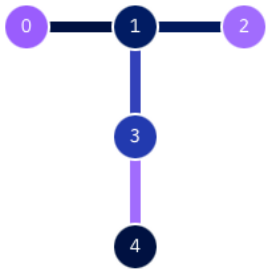

Figure S8: Connectivity representation of the IBMQ-Quito[12] device.

## Depolarizing Noise Model

Depolarizing noise channels are a common way to model decoherence and gate errors in quantum devices.[13,14] After observing that our two-qubit ($T$) gate-error is approximately an order of magnitude larger than our single-qubit (S) gate error, a noise model was constructed that provides $S$ as a linear function of $T$, $S = 0.1 * T$. This noise model was used to evaluate the effect of REM, corresponding to a range of depolarizing errors. The depolarizing noise was modelled in Qiskit as a probability $p'$ of applying each of the Pauli gates $X$, $Y$, or $Z$ after running a single-qubit gate, and as probability $p''$ of applying any combination of $P_i \otimes P_j$ where $P_i, P_j \in \{I, X, Y, Z\}$ after running a two-qubit gate.